\documentstyle[11pt,newpasp,epsf]{article}

\begin{document}

\title{New generations of stellar model atmospheres}

\author{Martin Asplund}
\affil{Research School of Astronomy and Astrophysics,
Mt Stromlo Observatory, Cotter Road, Weston, ACT 2611, Australia}

\begin{abstract}

Stellar model atmospheres form the basis for any element abundance
determination and hence are crucial ingredients
for studies of stellar, galactic and cosmic evolution.
With recent observational progress with the advent of
8m-class telescopes and efficient spectrographs, the dominant
source of uncertainty today originates with the assumptions and approximations
in the analyses, emphasizing the great need for continuing efforts in
improving the realism of stellar atmosphere modelling.
In the present contribution I will describe recent progress in this
regard by focussing on three complementary types of
model atmospheres: line-blanketed non-LTE models of hot stars,
3D hydrodynamical models of cool stars and
semi-empirical models for large-scale stellar abundance analyses.

\end{abstract}


\section{Introduction}

Determining stellar element abundances play an integral role in
most endeavours to improve our understanding of stellar, galactic
and cosmic evolution.
The term {\em observed abundances} is somewhat of
a misnomer, however, since the chemical composition of course can not
be inferred directly from an observed spectrum.
The obtained abundances are therefore never more trustworthy than the models employed
to analyse the observations.
The derivation of accurate abundances require realistic models
of both the spectrum formation region and the spectrum
formation process, an obvious fact which unfortunately is often overlooked
in the analysis and in assessing the systematic errors.
Costly new observational facilities like large telescopes and
sophisticated instrumentation will only reach their full potential once
combined with the appropriate analysis tools. Today, the uncertainties are no
longer dominated by observational errors but by shortcomings in the
modelling of stellar atmospheres and line formation.
Nevertheless, much progress has been made
lately in this respect and here I will highlight a few of these.
Unfortunately due to page limitations I am unable to
discuss many other impressive works for which I sincerely
apologize. I will concentrate on
one area each for hot and cool stars:
line-blanketed non-LTE modelling and 3D hydrodynamical treatment of convection,
respectively.
%Potential areas for future improvements will also be identified.
I will conclude the review by making a call to arms to develop non-LTE inversion
methods which are urgently needed to enable analysis of very large stellar samples.
%for studies of galaxy formation and evolution among other things
%using multi-object spectrographs and forthcoming satellite missions.

\section{Hot stars}

\subsection{Line-blanketed non-LTE model atmospheres}

With the high temperatures encountered in hot star atmospheres, radiative
rates mostly overwhelm collisional ones, rendering the assumption of local
thermodynamic equilibrium (LTE) unacceptable.
As a consequence the rate equations for all atomic level populations
(assuming time-independence: statistical equilibrium)
must be solved simultaneously with the radiative transfer equation,
as well as the hydrodynamical equations of conservation of mass, momentum
and energy (which in the absence of a wind simplify to hydrostatic and
radiative equilibria), in a plane-parallel or spherical 1D geometry.
Since the populations depend on the radiation field, which in turn is
determined by the opacities and hence populations, this system of equations is
extremely non-local and non-linear with everything in principle depending
on everything else, everywhere else. Complicating factors also arise from the fact that
the intense radiation field often drive massive stellar winds, which among other
things cause severe problems for the treatment of radiative transfer by
shifting spectral lines in and out of the shadows of other lines formed in deeper layers.

The last couple of decades has seen very impressive progress in constructing
non-LTE model atmospheres (e.g. Hillier \& Miller 1999; Pauldrach et al. 2001;
Werner 2003 and references therein). Much of this is due to the
development of efficient numerical algorithms such as the approximate
lambda iteration technique (e.g. Scharmer 1981). The mean intensity $J_\nu$ is
obtained from the source function $S_\nu$ using an iterative procedure
$J_\nu^n = \Lambda^*S_\nu^n + (\Lambda - \Lambda^*)S_\nu^{n-1}$
through an approximate (e.g. local) lambda operator $\Lambda^*$;
$\Lambda S_\nu^{n-1}$ simply denotes a formal radiative transfer solution with
the current estimate of $S_\nu$ (see e.g. Werner 2003 for a detailed description).
Improved convergence properties
can also be achieved through preconditioning of the equations.
Equally important for the success has been
the advent of appropriate atomic data and methods for the statistical
representation of them. Data from Opacity Project, Iron Project and the like
has allowed line-blanketing from millions of spectral lines
involving thousands of atomic levels from different species to be taken into account
with $\sim 10^5$ frequency points
using the superlevel concept (Anderson 1989).

Some work in this arena still remains however. Often the radiatively-driven wind is not
self-consistently calculated from the theoretical radiative acceleration
(the so-called wind-momentum problem of WR and similar stars)
and instead a parametrized mass loss rate and
velocity law are prescribed (e.g. Castor et al. 1975).
Alternatively the calculations are restricted to either the quasi-hydrostatic photosphere
or the supersonic wind parts.
Observationally there is increasing evidence for both time-dependence and
spatial inhomogeneities in the winds of hot stars in the form of shocks and clumps
being accelerated outwards (e.g. Eversberg et al. 1998).
Needless to say, time-dependent 3D non-LTE calculations is a very daunting task indeed but
given the enormous progress achieved lately it is not inconceivable that also these
issues will be addressed with the next generation of models in the coming years.

\subsection{Implications for CNO abundances}

The introduction of massive line-blanketing in the new generation of model
atmospheres of hot stars directly affects the atmospheric structure, the
ionization balance and the
predicted emergent flux distribution. Indirectly it also modifies the estimated
stellar parameters, in particular the effective temperature $T_{\rm eff}$ and the
mass loss rate. For main sequence O stars the inclusion of line-blanketing can lead
to a substantial decrease in $T_{\rm eff}$ of 1000-4000\,K (Martins et al. 2002) at solar
metallicity, which obviously also has an impact on the derived element abundances.
Even larger differences are expected for giants and supergiants.

C, N and O are among the relatively few elements for which abundances can be determined
in OB stars. With the improved atmospheric and ionization structures, the overall agreement
between observations and predictions is in general quite impressive, with the exception of some
lines formed in the extended wind (Pauldrach et al. 2001; Crowther et al. 2002).
A common denominator in most analyses of Galactic and SMC/LMC O supergiants is the
existence of non-solar CNO abundance ratios with a large N enrichment and weak C and O depletion.
This indicates mixing of unprocessed and CNO-processed material, possibly related to
rotational mixing (Meynet \& Maeder 2000).
Recently the first results for other external galaxies have appeared based on
8-10m telescopes, again suggesting He and N enrichments (Bresolin et al. 2002).

\section{Cool stars}

\subsection{3D time-dependent hydrodynamical model atmospheres}

For late-type stars such as the Sun, the most pressing uncertainty stems from
our poor understanding of convection. The visible surface from
which the stellar spectrum originates represents the transition from convective
to radiative energy transport.
% and thus any shortcoming in the modelling of
%this region will directly impact the conclusions inferred from the observed spectrum.
Traditionally convection has been treated by the mixing length theory (MLT)
or some close relative thereof. In addition to being a very rudimentary description,
MLT is a local theory,
which ignores convective overshoot and radiative transfer effects. A far more realistic
%but significantly more computationally demanding
approach is to solve the hydrodynamical equations coupled to the equation
of radiative transfer. Recently much progress has been made
in performing 3D radiative-hydrodynamical simulations of
stellar surface convection (e.g. Nordlund \& Dravins 1990; Freytag et al. 1996, 2002;
Stein \& Nordlund 1998; Asplund et al. 1999, 2000a;
Asplund \& Garc\'{\i}a P{\'e}rez 2001; Ludwig et al. 2002),
which subsequently can be applied to studies
of spectral line formation. These investigations have clearly shown that in many
cases standard 1D analyses are very misleading in terms of derived
element abundances.

The 3D model atmospheres developed by our group have been
computed with a 3D, time-dependent, compressible, explicit,
radiative-hydrodynamics code (Stein \& Nordlund 1998).
The hydrodynamical equations for conservation of mass, momentum
and energy are solved on a Eulerian mesh with gridsizes of $\approx 100^3$.
The physical dimensions of the grids are sufficiently large
to cover many granules simultaneously and extended enough in the vertical
direction to reach almost adiabatic conditions in the bottom. In terms of
continuum optical depth the simulations extend
at least up to log\,$\tau_{\rm Ross} \approx -5$, which for
most purposes are sufficient to avoid
numerical artifacts of the open upper boundary on spectral line formation.
Periodic horizontal boundary conditions are employed, allowing the simulations
to describe a small and representative region of the stellar surface (Fig. 1).
The temporal evolution of the simulations cover several convective turn-over
time-scales to allow thermal relaxation to
be established and to obtain statistically significant
average atmospheric structures and spectral line profiles following disk-integration.
The input parameters discriminating different models are the
surface gravity log\,$g$, metallicity [Fe/H] and the entropy of the
inflowing material at the bottom boundary. The effective temperature of
the simulation is therefore a property which depends on the
entropy structure and evolves with time
following changes in the granulation pattern.

\begin{figure}
\plotone{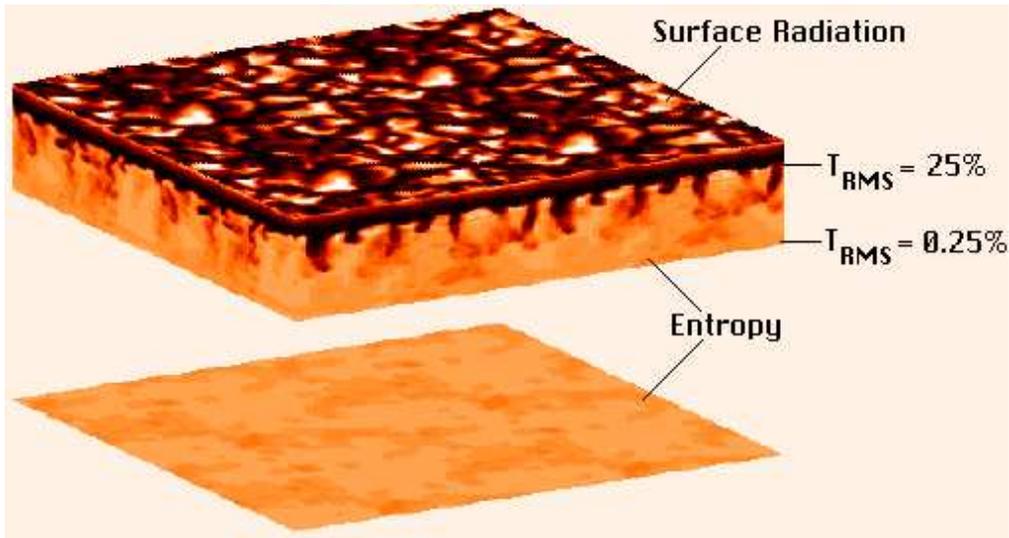}
\caption{A snapshot from a 3D hydrodynamical model atmosphere of the Sun depicting
the emergent continuous intensity on the top layer and the entropy structure on sides
and the displaced bottom layer. The maximum temperature contrast occurs close to
the visible surface, revealing the granulation pattern of warm upflows and cool
downdrafts.
%In deeper layers the convective flux can be carried by a smaller contrast
%due to the higher densitities.
}
\end{figure}

In order to obtain a realistic atmospheric structure,
it is crucial to have the best possible input physics, and
properly account for the energy exchange between
the radiation field and the gas.
The adopted equation-of-state is that of Mihalas et al. (1988),
which includes the effects of ionization, excitation
and dissociation. The continuous opacities come from
the Uppsala package (Gustafsson et al. 1975 and subsequent
updates) while the line opacities are from Kurucz (1993).
The 3D radiative transfer is solved at each time-step
%using 9 inclined rays (2 $\mu$-angles and 4 $\varphi$-angles, plus the vertical)
under the assumptions of local
thermodynamic equilibrium (LTE, $S_\lambda = B_\lambda$)
and opacity binning.
% (Nordlund 1982).
%The four opacity bins are designed to correspond to
%continuum and weak, intermediate and strong lines.
The opacity binning includes the effects of line-blanketing
in a manner reminiscent of opacity distribution functions.
%In order to improve the numerical accuracy, the radiative
%transfer is solved on a finer grid and the results subsequently
%interpolated back to the original grid for the hydrodynamical variables
%at each timestep.

It should be stressed that the here described 3D hydrodynamical model
atmospheres do not include any free parameters which are tuned to
improve the agreement with observations. Yet the simulations are highly
successful in reproducing key observational constraints, such as
helioseismology, detailed spectral line shapes,
and granulation characteristics (e.g. Stein \& Nordlund 1998;
%Rosenthal et al. 1999;
Asplund et al. 2000a,b).
Thus the usual free parameters (mixing length parameters, micro- and
macroturbulence) hampering classical 1D analyses have finally become
obsolete with the advent of the new generation of 3D model atmospheres.
The major drawbacks with such 3D modelling is their computationally
demanding nature and that sofar only a relatively limited stellar parameter
space has been chartered.
Through the continuous improvement in computer hardware and software the
former problem will be steadily relaxed in the coming years, while already
work is being undertaken to extend the modelling to other types of stars, most
notably A stars (Freytag et al. 1996), M dwarfs (Ludwig et al. 2002),
supergiants (Freytag et al. 2002) and red giants.

\subsection{Implications for CNO abundances}

Spectrum formation is a highly non-local and non-linear process.
It is therefore not surprising that 3D analyses in general yield element abundances
which are not the same as with traditional 1D investigations, given both
the different photospheric mean structures and the presence
of spatial inhomogeneities.
%In many cases the differences are not alarming
%and often negligible compared with other uncertainties such as non-LTE effects.
In many cases the differences can be profound with far-reaching consequences.
%Of course only with the results of
%the more detailed calculations in hand can the systematic errors in standard analyses
%be evaluated.
Here I will limit the discussion on the impact of 3D model atmospheres
on C, N and O for the Sun and metal-poor stars.

{\bf Sun:} The solar spectrum offers a multitude of atomic and molecular lines to
determine the abundances of C, N and O. Primary diagnostics are:
[C\,{\sc i}] 872.7\,nm, high-excitation C\,{\sc i} lines, and
CH vibration-rotation transitions in the IR for C; permitted N\,{\sc i} lines
and NH vibration-rotation lines in the IR for N;
[O\,{\sc i}] 630.0\,nm, high-excitation O\,{\sc i} lines,
and OH vibration-rotation and rotation-rotation transitions in the IR for O.
Secondary, supportive roles are offered by electronic CH, C$_2$, CN and OH lines.
Astonishingly, with 1D model atmospheres the discrepancies between different indicators
amount to as much as 0.2-0.3\,dex, casting doubts on the use of such models for
stars where not all types of lines can be employed. In particular for O the dissonance
is striking with the often preferred OH lines
(e.g. Grevesse \& Sauval 1998)
indicating about 0.3\,dex higher abundance than the O\,{\sc i} 777\,nm triplet with the
[O\,{\sc i}] line in between.

\begin{figure}
\plotone{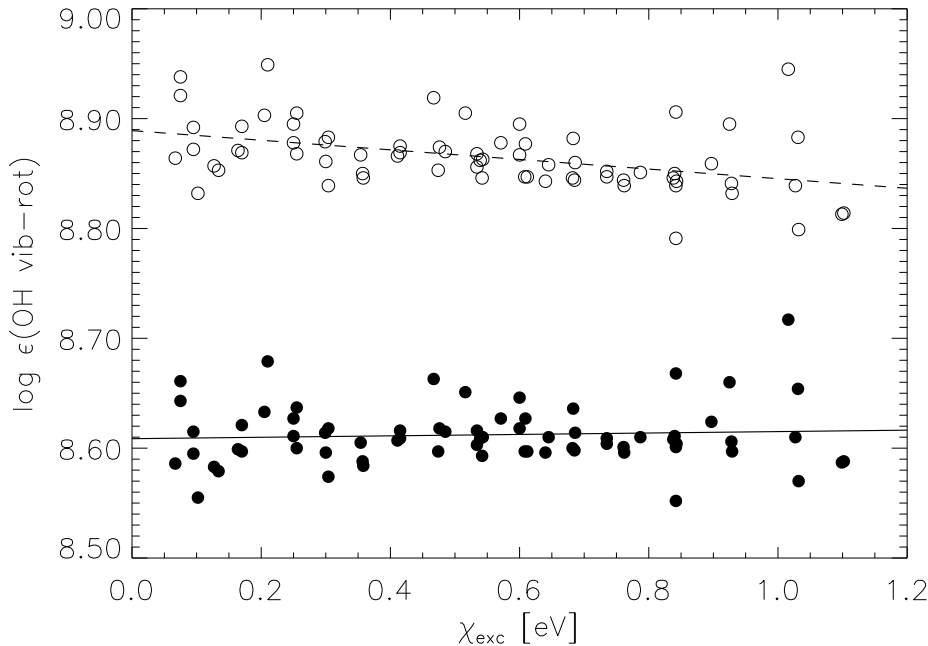}
\caption{The derived solar O abundances from the OH vibration-rotation lines in
the IR using a 3D model atmosphere (filled circles) and the semi-empirical 1D
Holweger-M\"uller model (open circles) as a function of excitation potential of
the lower level.
%The lines denote the least square fits for the two cases.
}
\end{figure}

We have recently re-analysed all of the above-mentioned solar diagnostics using
our 3D hydrodynamical solar model atmosphere, the best possible atomic and molecular data,
and high-quality solar atlases.
Reassuringly, we find excellent agreement between different lines for
C, N and O, in sharp contrast to the 1D case. In particular, the presence of cool
pockets in the photosphere due to convection produces stronger molecular lines and
hence lower derived abundances compared with standard 1D model atmospheres.
Our preliminary results indicate log\,$\epsilon(\rm C) = 8.41 \pm 0.03$,
log\,$\epsilon(\rm N) = 7.80 \pm 0.04$, and
log\,$\epsilon(\rm C) = 8.66 \pm 0.03$, where the here quoted uncertainties are only
the standard deviations between the different abundance indicators
(Allende Prieto et al. 2001, 2002; Asplund et al., in preparation).
It is noteworthy that this represents a substantial decrease from previously
adopted values (e.g. Fig. 2) but is in good agreement with
results for nearby B stars, local interstellar medium and solar corona/wind
measurements.
The resolution of these long-standing discrepancies gives further credibility to
the new generation of 3D hydrodynamical model atmospheres.

{\bf Metal-poor stars:} The differences between 1D and 3D model atmosphere
predictions are perhaps most pronounced in metal-poor halo stars (Asplund et al. 1999).
Due to the weak coupling through spectral lines between the radiation field and
the gas at low metallicities, the temperatures in the optically thin photospheric layers
can reach very low values with a proper treatment of convection, significantly below
the radiative equilibrium temperatures enforced in standard 1D model atmospheres.
For spectral lines sensitive to those atmospheric depths, the resulting differences
can be profound. In particular, molecular features, minority ionization stages and
low-excitation lines are very vulnerable and significant systematic errors can be
expected in 1D analyses. This inadequate atmospheric modelling probably lies at
the heart of the long-standing oxygen problem in metal-poor stars, i.e. that
different diagnostics support either a flat plateau at [O/Fe]\,$\sim +0.4$ (not accounting
for the new low solar O abundance mentioned above) or linear rising trend reaching
[O/Fe]\,$\sim +1$ at [Fe/H]\,$=-3$. In particular, the application of 3D model atmospheres
bring down the very high oxygen abundances from OH lines
(Asplund \& Garc\'{\i}a P{\'e}rez 2001) but also those of the [O\,{\sc i}] line
(Nissen et al. 2002); 3D non-LTE calculations of O\,{\sc i} yield similar results
as in 1D non-LTE (Asplund et al., in preparation). The final verdict is not yet in,
partly due other out-standing factors such as non-LTE effects on OH, but it
appears that the truth, not surprisingly perhaps, lies in between the hotly contested
plateau and linear trend (Nissen et al. 2002). Similarly, many conclusions about
C and N abundances in metal-poors probably have to be revised. Normally such
determinations are based on molecular lines such as CH, NH and CN, and hence the
abundances are likely significantly over-estimated in 1D. Naturally this would bring down
the claimed [C/Fe] and [N/Fe] trends at low metallicity, as well as the incidence of very
C-rich halo stars. Work is currently under way to address these issues.

\section{Large-scale stellar abundance analyses}

\subsection{The need for semi-empirical model atmospheres}

The new generations of stellar model atmospheres described above are both
theoretical, as are almost all other models.
However, the properties of the stellar atmospheres are encoded in the
observed stellar spectra which at least in principle can
be deciphered to yield the photospheric structure
%simultaneously with the element abundances
using inversion techniques.
This method has often been employed among solar physicists with
the Holweger-M\"uller (1974) and the
VAL-3C (Vernazza et al. 1981) solar models being the most
well-known examples. Surprisingly, such semi-empirical model atmospheres have
rarely been used for stellar spectroscopy.
I believe that the time is now ripe for pursuing such modelling,
in fact crucial in order to make optimum use of future
satellite and ground-based telescope facilities.

Perhaps the biggest problem with using the above-mentioned theoretical model atmospheres for
stellar abundance analyses is their computationally intensive nature.
%which means that the analyses are restricted to relatively few stars.
Although less sophisticated model atmospheres are
less CPU-demanding, also standard abundance analyses are time-consuming in
terms of the work involved. Even today the largest studies of stellar abundances
involve only a few hundred stars (e.g. Reddy et al. 2003).
To fully disentangle the formation and evolution of for example our
own Galaxy would require a far more ambitious approach with much
larger stellar samples (Freeman \& Bland-Hawthorn 2002).
Fortunately this is now becoming feasible with new multi-object
high-resolution spectrographs like FLAMES/VLT. Several other, even more
ambitious instruments are currently in the design phase or being
discussed in detail, such as the RAVE project
(Steinmetz et al. 2002) and the GAIA satellite
(Perryman et al. 2001).
The wealth of information these facilities will provide is truly
astonishing, as is the magnitude of data needed to be analysed.
Without an efficient automatic procedure to derive the inherent information in the
spectra, these projects will not come to an optimal execution.
Semi-empirical model atmospheres represent one such promising possibility, provided
the spectral resolving power is sufficient.

\subsection{Ingredients and outlook for the future}

Semi-empirical model atmospheres are obtained through an iterative procedure
aimed at improving the overall agreement
between the calculated and observed spectra. Since the temperature structure
is modified in this way the model does not in general fulfill flux constancy as
in standard theoretical model atmospheres, nor is there any need to invoke
uncertain and questionable recipes for convective energy transport, such as
the mixing length theory, which can lead to grossly misleading results
(e.g. Asplund et al. 1999; Asplund \& Garc\'{\i}a P{\'e}rez 2001).
In fact, this allows at least in principle a possibility
to gain insight to convection and non-standard energy transport, such as
acoustic and magnetic heating. With the temperature determined from the
observations, the gas pressure is normally fixed by the assumption of
hydrostatic equilibrium while the electron pressure is the result of
the ionization balance using the current estimate of the chemical composition.
Thus knowledge of the surface gravity is necessary from for example parallaxes or
photometry unless it can be obtained from satisfying the ionization balance of
some key elements.
Essentially all existing modelling is based on a 1D plane-parallel geometry but some
attempts with multi-component atmospheres have been made, corresponding for
example to convective up- and downflows (Frutiger et al. 2000).
Naturally, inversions must rely on accurate atomic data to yield realistic results.

The main disadvantages with semi-empirical modelling comes with the numerical
difficulties often associated with ill-posed inversion problems and the assumption
of LTE normally made in the line formation calculations. The former can to a large
extent be
overcome by including a wide range of spectral lines with different line formation
depths, ionization stages and excitation potentials. Of particular importance,
at least for late-type stars, is to include elements with both neutral
and ionized lines and strong lines with pressure-damped wings. Although not
yet utilized, hydrogen lines as well as
various molecular lines have a great potential in this respect.
The second major shortcoming can only be resolved by relaxing the assumption of LTE, which
automatically makes the inversion procedure much more challenging and cumbersome.
Very little work with the exception of the pioneering contribution
by Soccas-Navarro et al. (1998)
has been done on non-LTE inversion. It is, however, of absolute pivotal importance
to include such effects, given the fact that LTE is often a poor
assumption for the line formation process, leading otherwise to
erroneous atmospheric structures and inferred conclusions.
It suffices here to mention the case of Fe\,{\sc i} which offers the most
useful lines for inversion modelling in late-type stars but suffers from over-ionization,
in particular at low metallicities.

Recent endeavours like the Opacity Project and Iron Project have come a long way
in addressing the great needs for atomic data for non-LTE calculations.
Although it would be advantageous to treat all elements in non-LTE for the
inversions, as a first step it may be sufficient to do so for the most important
elements for determining the actual atmospheric structure, such as Fe, Mg and Ca,
while the line formation of
other significant species for late-type stars like H and various
molecules (e.g. CO) can likely be well approximated by LTE.
Other trace elements which are unimportant for setting the atmospheric structure
but whose abundance may be of interest can then be handled either in LTE or non-LTE
after the initial inversion has converged.
This procedure should restrict the computing time to just a few times
a normal non-LTE calculation with a fixed 1D atmosphere.
Adopting the superlevel concept to reduce the number of atomic levels for
complex atoms like Fe should speed up the computations further and bring
the non-LTE inversion concept fully within the realms of possibility for the coming
few years. This should enable truly grandiose studies of galaxy formation and evolution
possible with current and future observational capabilities.

\acknowledgments

I am grateful to the SOC and LOC for organizing an extremely stimulating
conference in honour of Prof. Andr\'e Maeder, for providing me with a
travel grant to attend the meeting and for showing great patience in waiting for
my contribution to the proceedings. I am already looking forward to
the conference celebrating Andr\'e's 70th birthday, which will hopefully
take place in a similarly wonderful environment and atmosphere.
Even further into the future I hope I am not too old
to also participate in the conference
held in connection with Andr\'e's retirement.

\end{document}